\documentclass[twoside,leqno,twocolumn]{article}  
\usepackage{ltexpprt} 
\usepackage{graphicx}
\usepackage{balance} 
\usepackage{color}
\usepackage{dsfont}
\usepackage{amsfonts}
\usepackage{amsmath}
\usepackage{subcaption}
\usepackage{times}
\usepackage{epsfig}
\usepackage{algorithm} 
\usepackage{algorithmic}
\usepackage{tabularx}
\usepackage{multirow}

\newcolumntype{I}{!{\vrule width 3pt}}
\newlength\savedwidth

\newlength\savewidth
\newcommand\shline{\noalign{\global\savewidth\arrayrulewidth
                            \global\arrayrulewidth 1.5pt}%
                   \hline
                   \noalign{\global\arrayrulewidth\savewidth}}

\begin{document}

\title{Decentralized Recommender Systems}
\author{Zhangyang Wang\thanks{\scriptsize Beckman Institute, University of Illinois at Urbana-Champaign, IL 61821. \{zwang119, xliu102, chang87, t-huang1\}@illinois.edu.}\and Xianming Liu$^*$\and Shiyu Chang$^*$\and Jiayu Zhou\thanks{\scriptsize Samsung Research America, San Jose, CA 95134. jiayu.zhou@samsung.com} \and Guo-Jun Qi\thanks{\scriptsize University of Central Florida, Orlando, FL, 32816, USA. guojun.qi@ucf.edu.}  \and Thomas S. Huang$^*$ \\}
\date{}

\maketitle


\begin{abstract} \small\baselineskip=9pt 
This paper proposes a \textit{decentralized recommender system} by formulating the popular collaborative filleting (CF) model into a decentralized matrix completion form over a set of users. In such a way, data storages and computations are fully distributed. Each user could exchange limited information with its local neighborhood, and thus it avoids the centralized fusion. Advantages of the proposed system include a protection on user privacy, as well as better scalability and robustness. We compare our proposed algorithm with several state-of-the-art algorithms on the FlickerUserFavor dataset, and demonstrate that the decentralized algorithm can gain a competitive performance to others.
\end{abstract}

\section{Introduction}

The paper discusses the decentralized recommender systems, which is in contrast to the typical recommender systems built on centralized infrastructures (the ``cloud'', etc. ). The decentralized network \cite{CY} had been thoroughly investigated in control and communication fields, defined as a set of distributed but connected agents, who are generally not strongly connected in a graph theoretic sense.  Each agent collects data by itself, and executes computation via limited communication within only local neighborhoods. Fig. 1 shows a comparison of centralized versus decentralized network structures. Specifically, in a decentralized recommender system, individual users / user-groups can be viewed as network agents. Each user holds his or her own ratings as partially observed data. The data cannot be accessed by any intermediate point or centralized server in the network. Therefore, it has a potential effect on protecting user data privacy against both the cloud server and some malicious eavesdropping over uploading channels. For a large-scale network of mobile users, the decentralized models own a better scalability since users are only locally connected. Since data storages and computations are fully distributed, the decentralized systems also become robust to center (cloud) or individual agent (user) failures. 
\begin{figure}[htbp]
\centering
\begin{minipage}{0.22\textwidth}
\centering {
\includegraphics[width=\textwidth]{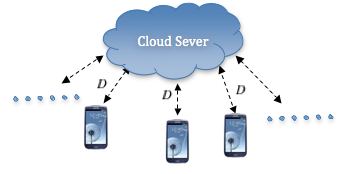}
}\end{minipage}
\begin{minipage}{0.22\textwidth}
\centering{
\includegraphics[width=\textwidth]{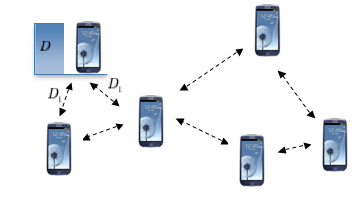}
}\end{minipage}
\caption{The comparison of a centralized network (left) and a decentralized network (right).}
\label{fig:basic}
\end{figure}

To our best knowledge, we are the first studying and designing a decentralized recommender system. There have been some recent interests in investigating algorithms in a decentralized fashion \cite{DMC, Dl1}, and moreover,  preliminary literatures to reveal the convergence \cite{DADMM} and dynamics \cite{ALM} properties. All above make it solid and promising to build application scenes on a decentralized network structure.

\section{Model and Algorithm}

As a most popular tool in recommender system, Collaborative Filtering (CF) is usually formulated as matrix factorizations problem~\cite{9} \cite{chang2014factorized}. It predicts user rating $\mathbf{R}_{i,j}$ (of $i$-th user on $j$-th item) as a dot product of the user profile of the $i$-th user, denoted as a row vector $\mathbf{U}^i$, and the item profile of the j-th item denoted as a column vector  $\mathbf{V}_j$, i.e., $\mathbf{R}_{i,j} = \mathbf{U}^i \mathbf{V}_j$. The recommendation problem can be formulated as solving the following matrix factorization problem:
\begin{equation}
\begin{array}{l}\label{MC}
\min_\mathbf{U,V,Z} \frac{1}{2}||\mathbf{UV-Z}||_2^2  \\
s.t. \quad P_{\Omega} (\mathbf{Z}) = P_{\Omega} (\mathbf{R})
\end{array}
\end{equation}
Here $P_{\Omega}$ denotes the projection over the set of available ratings, and $\mathbf{Z}$ is an auxiliary matrix.

\begin{algorithm}[ht]
\caption{Decentralized matrix completion (DMC) algorithm for solving (\ref{MC})}
\begin{algorithmic}[1]
\REQUIRE $P_\Omega(\mathbf{R}_i)$, ($i$ = $1,2,...,L$); initializations of $\mathbf{U}_i$, $\mathbf{V}_i$, and $\mathbf{Z}_i$ for each $i$-th user ($i$ = $1,2,...,L$); step size $\beta$; ITER

\STATE FOR t=1 to ITER DO

\STATE Each $i$-th user updates $\mathbf{V}_i$:  $\mathbf{V}_i =\mathbf{ (U_i^TU_i)^{-1}U_i^TZ_i}$

\STATE  Each $i$-th user updates $\mathbf{Z}_i$:  $\mathbf{Z}_i = \mathbf{U_iV_i}+P_\Omega(\mathbf{R_i - U_iV_i})$

\STATE Each $i$-th user propagates $\mathbf{U}_i$ to its one-hop neighborhood $N_i$.

\STATE Each $i$-th user updates $\mathbf{U}_i$: \\
$\mathbf{U}_i = \frac{\mathbf{Z}_i \mathbf{V}_i^T -  \mathbf{a}_i + \beta \sum_{j \in N_i}  \mathbf{U}_j}{1+ 2 \beta |N_i|}$\\
$ \mathbf{a}_i =  \mathbf{a}_i  + \beta (|N_i|\mathbf{U}_i -  \sum_{j \in N_i}  \mathbf{U}_j$

\STATE END

\ENSURE  $\mathbf{U}_i$, $\mathbf{V}_i$, and $\mathbf{Z}_i$, $i$ = $1,2,...,L$
\end{algorithmic}
\end{algorithm}

It is assumed that CF is performed by $L$ users jointly in a decentralized manner, and $\mathbf{R}$ is segmented into $L$ non-overlapped parts, denoted as $\mathbf{R}_i$, $i$ = $1,2,...,L$. For example, the easiest case to segment $\mathbf{R}$ is to divide by columns. 
The $i$-th user ($i$ = $1,2,...,L$) observes $\mathbf{R}_i$. Note some level of synchronization is still required to collaboratively utilize information from all users. The trade-off strategy is \textbf{to share partial data only among users in the local neighborhood}.  After observing the problem structure, authors in \cite{DMC} suggested an variant of nonlinear Gauss-Seidel (GS) iterations, named \textit{decentralized matrix completion} (\textbf{DMC}) algorithm. The $i$-th user will hold $\mathbf{R}_i$, as well as $\mathbf{U}_i$, $\mathbf{V}_i$, and $\mathbf{Z}_i$ based on its own computations. Note $\mathbf{U}_i$ here is of the same size as $\mathbf{U}$, and $\mathbf{Z}_i$ of the same dimension as $\mathbf{R}_i$, so in other words, $\mathbf{Z}_i = \mathbf{U}_i \mathbf{V}_i$. In each iteration, the $i$-th user first updates $\mathbf{V}_i$ and $\mathbf{Z}_i$ independently, then exchanging $\mathbf{U}_i$ with its one-hop connected neighborhood users,  and finally update $\mathbf{U}_i$ via the average consensus algorithm\cite{Con}. The algorithm is summarized in Algorithm I. It obtains similar reconstruction errors, compared to centralized solutions \cite{DMC}.

\section{Experiments}

We compare our proposed algorithm with state-of-the-arts in this section, including Probabilistic Matrix Factorization (\emph{PMF}) \cite{pmf08}, and Collaborative Topic Modeling (\emph{CTR}) \cite{Chong11Collaborative}, on a collected image recommendation dataset from Flickr. 

\textbf{FlickrUserFavor dataset}: The dataset contains $350,000$ images collected from Flickr, from 140 user groups and uploaded by 20,298 users. We use the {\em ``like''} feedback provided by users as binary ratings. $75\%$ of the rating matrix is used as training, and $25\%$ as testing.


\textbf{Evaluation Measurement}: We use the averaged ranked order of all the rated images in the testing dataset for a specific user to evaluate performances. Among these ranked images, we determine those for which a user has exhibited a {\em ``like''} preference in the {\em test data}, and report the {\em average percentile score} (APS) of the ranked images which are indeed preferred by the user. The lower the APS, the better the algorithm is, which means the user preferred images are ranked in top positions. Finally, the {\em mAPS} is reported with the mean of the APS scores for all target users.

\textbf{Performance Comparision}: PMF \cite{pmf08} is the most classical collaborative filtering algorithm for recommender system. Wang et.al. also proposed the Collaborative Topic Model \cite{Chong11Collaborative} which involves both content and user ratings, where we use the Hierarchical Gaussianization (HG) \cite{zhou2009hierarchical} as the image features. For the proposed decentralized algorithm, we set the rank as 64, and using 8 agents. Detailed \emph{mAPS} comparisons are shown in Table~\ref{tab:performance}. It is shown that DMC is capable to achieve competitive performance as \emph{CTR}, while is far better than \emph{PMF}. Moreover, we do not use any content information in our algorithm, while \emph{CTR} uses content to indicates similarities between items. That suggests a further direction improve our algorithm too.

\begin{table}[htb]
\centering
\caption{Performances of the proposed approaches compared with other baseline methods. The second column indicates whether or not the algorithm uses content information. Without using content information and the fusion center, the proposed algorithm achieves a competitive performance.}
\label{tab:performance}
\begin{tabularx}{0.8\linewidth}{X|X|X}
\shline
\emph{method}       				& Content		&\emph{mAPS}              \\
\hline
PMF \cite{pmf08}    				& N 			& 61.99             \\
\hline
CTR \cite{Chong11Collaborative}      	& Y 			& 52.76                \\
\hline
DMC							& N 			& 53.46		\\
\shline
\end{tabularx}
\end{table}

\section{Conclusion}
This paper discusses a decentralized recommender system. We formulate the popular collaborative filleting model into a decentralized matrix completion problem. Each user, with only partial rating data, can exchange user profile factors with its local neighborhood, while keep item profile factors private. We compare our proposed algorithm with several state-of-the-arts on the FlickerUserFavor dataset, and illustrate comparable results to the conventional ones.

\section{Acknowledgement}
The work is supported by Samsung under the 2013 GRO project "Device centric Social Network Platform".


\end{document}